\documentclass[a4paper,11pt]{article}
\usepackage{jinstpub} 
\usepackage{siunitx}
\usepackage[version=4]{mhchem}
\usepackage{amsmath}
\usepackage{tikz}
\usetikzlibrary{positioning}

\title{Performance Reconstruction of Eco-Friendly Gas Mixtures for Improved Resistive Plate Chambers at GIF++ Using Geant4}

\author[a,1]{V.O. Ramirez-Beltran\note{Corresponding author.}}
\author[b]{Cecilia Uribe Estrada}
\author[a]{Mauricio Flores Geronimo}
\author[c]{F. Lagarde}

\affiliation[a]{Universidad Iberoamericana,\\
Mexico City, Mexico}

\affiliation[b]{Benem\'erita Universidad Aut\'onoma de Puebla,\\
Puebla, Mexico}

\affiliation[c]{University of Science and Technology of China}

\emailAdd{vicolenin.ramiez@cern.ch}

\abstract{
A macroscopic reconstruction is developed to infer iRPC performance using Geant4 observables and one experimental anchor. The Geant4 energy deposition is used to estimate the primary ionization yield, while the efficiency turn-on is modeled through an induced-charge description encoded in an effective gain $G(E)$. The absolute scale is fixed by calibrating the standard CMS mixture to its GIF++ efficiency curve and extracting macroscopic Townsend parameters $(A,B)$. The same procedure is propagated to four alternative mixtures, including two HFO and CO$_2$ eco-friendly blends, to reconstruct efficiency curves and working points, enabling detector mixture screening without microscopic transport inputs.
}

\keywords{
iRPC detectors;
Eco-friendly gases;
Geant4 simulations;
Townsend coefficient reconstruction
}

\arxivnumber{} 

\begin{document}
\maketitle
\flushbottom

\section{Introduction}\label{sec:intro}

The Compact Muon Solenoid (CMS) experiment at the Large Hadron Collider (LHC) employs a multilayer muon detection system based on three complementary technologies: Drift Tubes (DT), Cathode Strip Chambers (CSC), and Resistive Plate Chambers (RPC) \cite{pinheiro2024ichep}. Within this system, the iRPCs extend the muon coverage into the forward endcap regions and contribute to trigger redundancy and timing performance. They are designed to operate efficiently under the high-rate and high-background conditions expected during the High-Luminosity LHC (HL-LHC) phase \cite{lopez25th}.

The operational behavior of the iRPCs under HL-LHC background conditions is studied through beam tests at the Gamma Irradiation Facility (GIF++) at CERN \cite{lopez25th}. The GIF++ provides a controlled mixed-field environment that allows systematic measurements of the iRPC efficiency, cluster-size distributions, and time-resolution stability as functions of the applied voltage and irradiation level \cite{gomes2024performance}. These measurements constitute the official reference data used by the CMS collaboration to define the operational working points and stability criteria of the muon system \cite{lopez25th}.

In this work, a Monte Carlo simulation framework based on Geant4 is developed to model the GIF++ mixed radiation environment and its interaction with iRPC detectors. The simulation reproduces the combined gamma and muon irradiation conditions used in beam tests and is used to reconstruct detector performance, including efficiency and gain, as functions of the applied voltage and irradiation level. Experimental reference measurements from GIF++ are used to feed the simulation and generate the detector response, allowing a direct comparison between reconstructed and measured observables to assess the capability of the model to describe the observed detector behavior. The simulation is applied to five gas mixtures, including three previously studied mixtures and two eco-friendly mixtures currently under experimental study.

\section{Experimental arrangement}\label{sec:experimentalArrangement}

\begin{figure}[htbp]
\centering
\includegraphics[width=0.5\textwidth]{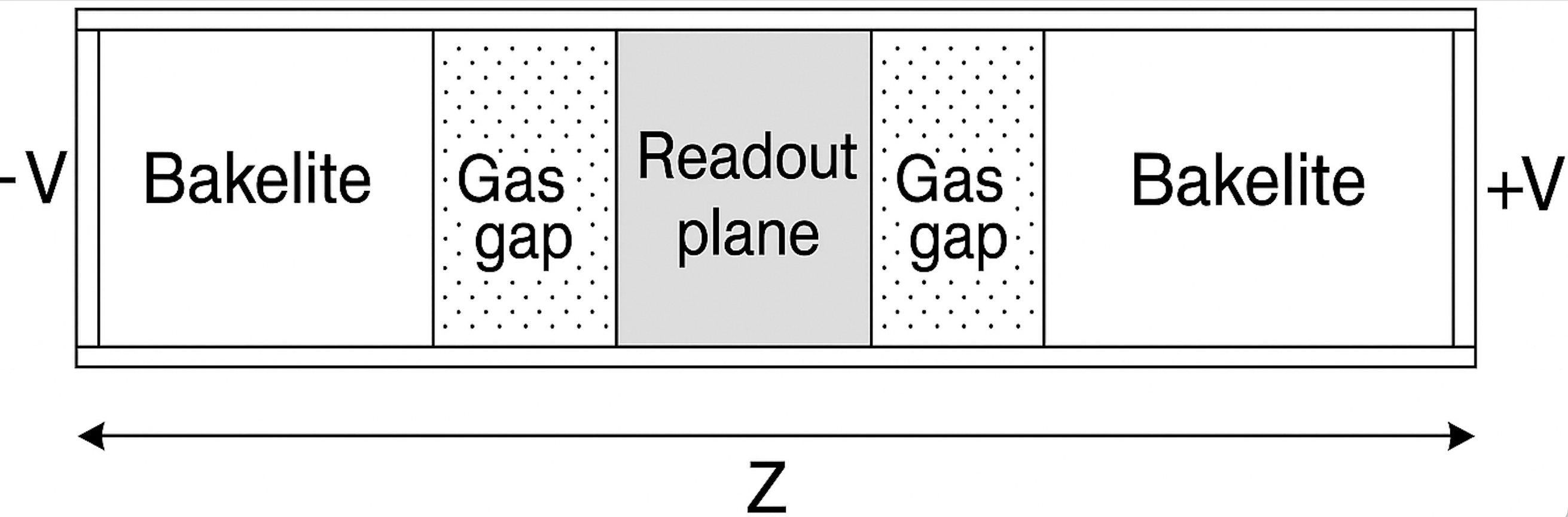}
\caption{Schematic geometry of the double gap iRPC simulated in this work.\label{fig:RPC2D}}
\end{figure}

The detector simulated in this work and implemented in the Geant4 framework corresponds to a double gap improved Resistive Plate Chamber (iRPC) with a central readout plane, shown schematically in Fig.~\ref{fig:RPC2D}. Each gas gap has a thickness of \SI{1.4}{mm}, matching the geometry of CMS iRPC prototypes currently tested at GIF++. The coordinate system and electrode positions defined in Fig.~\ref{fig:RPC2D} are used throughout the field calculations and efficiency reconstruction.

All materials, densities, and layer thicknesses reproduce those used in CMS iRPCs. The active medium is a gaseous mixture confined between two high-pressure laminate (HPL) electrodes  with relative permittivity $\varepsilon_r \approx 10$, while the copper readout strips are located at the mid-plane with an effective thickness of \SI{35}{\micro\meter}. Outer acrylic and aluminum layers are included to reproduce realistic boundary conditions and secondary interactions found in experimental beam-test configurations.

\subsection{Gas mixtures}

Five gas mixtures are considered in this study. The standard CMS mixture (STD) is used as a reference configuration, as it has been extensively studied in RPC and iRPC detectors under GIF++ conditions \cite{pinheiro2024ichep}. Two CO$_2$-based mixtures (MIX~I and MIX~II), previously investigated in RPC test programs, are included as intermediate reference cases \cite{supratik2025perspectives}. The remaining two mixtures, ECO1 and ECO2, are HFO1234ze/CO$_2$-based eco-friendly candidates that have been investigated in recent iRPC studies at GIF++ \cite{gomes2024performance}.

The compositions of the five gas mixtures considered in this study are summarized in Table~\ref{tab:mixtures}. All mixtures are defined at a temperature of $T = 293.15$~K and a pressure of $p = 0.957316$~atm. The table reports the volumetric fractions of each gas component for the standard CMS mixture (STD), the CO$_2$-based mixtures MIX~I and MIX~II, and the HFO1234ze/CO$_2$ eco-friendly mixtures ECO1 and ECO2.

\begin{table}[htbp]
\centering
\caption{Gas mixtures used in this study. Compositions are given in volume percent at $T=293.15$~K and $p=0.957316$~atm.}
\label{tab:mixtures}
\smallskip
\smallskip
\begin{tabular}{l|ccccccc}
Mixture & HFO & C2H2F4 & CO2 & Ar & N2 & iC4H10 & SF6 \\
\hline
STD      & 0  & 95.2 & 0  & 0 & 0    & 4.5 & 0.3 \\
MIX~I    & 0  & 0    & 60 & 5 & 35   & 0   & 0   \\
MIX~II   & 0  & 0    & 60 & 5 & 34.5 & 0   & 0.5 \\
ECO1     & 35 & 0    & 60 & 0 & 0    & 4   & 1   \\
ECO2     & 25 & 0    & 69 & 0 & 0    & 5   & 1   \\
\end{tabular}
\end{table}

The standard and CO$_2$-based mixtures (STD, MIX~I, and MIX~II) are used as reference cases to anchor the macroscopic Geant4 framework to experimentally established detector behavior. The eco-friendly mixtures ECO1 and ECO2 constitute the primary focus of this work, for which the efficiency and gain reconstruction methodology presented in the following sections is applied and discussed.

\subsection{Reproduction of the GIF++ irradiation environment}

At GIF++, detector performance is evaluated under the simultaneous exposure to a high-energy muon beam and a photon background generated by a $^{137}$Cs source~\cite{pfeiffer2017radiation}. In this work, these two components are reproduced by considering a muon beam with a kinetic energy of \SI{6}{GeV} and a photon field composed of \SI{662}{keV} gamma rays, corresponding to the characteristic emission of the $^{137}$Cs source.

The photon background is modeled using a $^{137}$Cs source emitting \SI{662}{keV} gamma rays. The number of photons generated per event is derived from a target surface flux equivalent to the GIF++ irradiation level.

\begin{equation} \label{eq:Ngamma_target}
N_\gamma^{\text{target}} = \Phi_\gamma \, A \, \Delta t ,
\end{equation}

where $\Phi_\gamma$ is the photon surface flux at the detector, $A$ the irradiated area, and $\Delta t$ the effective time window per event. For numerical stability, the photon multiplicity is limited to $N_\gamma \le 100$ per event.  
\section{Methodology} \label{sec:Methodology}

The purpose of this section is to define a fully macroscopic procedure to reconstruct the efficiency turn on of an improved Resistive Plate Chamber using a simulation framework. The approach relies exclusively on quantities directly provided by Geant4, namely primary ionization, energy deposition, and event level hit information, together with an experimental reference. In this context, detector efficiency is treated as a probabilistic quantity. For a given electric field configuration, efficiency is defined as the probability that an incident particle produces a detectable signal, rather than as a deterministic function of the deposited energy. This probability is quantified through the detector sensitivity, defined as

\begin{equation}
S(E) = \frac{N_{\mathrm{HIT}}(E)}{N_{\mathrm{tot}}(E)},
\label{eq:S_definition}
\end{equation}

where $N_{\mathrm{HIT}}(E)$ is the number of detected events at a given electric field $E$, and $N_{\mathrm{tot}}(E)$ is the total number of incident particles. This definition provides the starting point for reconstructing the field dependent detector response without invoking microscopic avalanche dynamics.

Each gas mixture and electric field configuration, the primary ionization yield is obtained directly from the energy deposited in the gas gap as simulated by Geant4.

For minimum ionizing muons, the energy loss in the gas is dominated by ionization processes and is essentially independent of the applied electric field over the operating range of RPC detectors. The mean number of primary electron ion pairs produced per event is therefore computed as

\begin{equation}
N_0(E) = \frac{\langle E_{\mathrm{dep}}(E) \rangle}{W_{\mathrm{mix}}},
\label{eq:N0_from_Edep}
\end{equation}

where $\langle E_{\mathrm{dep}}(E) \rangle$ is the average energy deposited in the gas gap at a given electric field $E$, and $W_{\mathrm{mix}}$ is the effective ionization energy of the gas mixture. The effective $W$ value depends only on the mixture composition and is computed as the average of the ionization energies of the individual gas components, weighted by their volume fractions.

\begin{equation}
W_{\mathrm{mix}} = \sum_i f_i\, W_i ,
\label{eq:Wmix}
\end{equation}

where $f_i$ is the volume fraction of component $i$ and $W_i$ is its mean ionization energy. The $W_i$ values used in this work are taken from standard references on gaseous detectors and electron ion pair formation.

\subsection*{Induced charge and field dependent response}

The connection between the field independent primary ionization yield and the detector sensitivity is established through the induced charge on the readout electrodes. At the macroscopic level, the signal formation in RPC detectors can be described using the Shockley--Ramo theorem. For a single drifting charge carrier, the induced charge on a given electrode is given by:

\begin{equation}
Q_{\mathrm{ind}} = q \left[ \phi_w(\mathrm{final}) - \phi_w(\mathrm{initial}) \right]
= q\, \Delta \phi_w ,
\label{eq:SR_basic}
\end{equation}

where $\phi_w$ is the weighting potential associated with the electrode geometry. This expression defines a purely geometrical coupling factor, Eq.~(\ref{eq:fgeom}), which accounts for the detector geometry and readout configuration,

\begin{equation}
f_{\mathrm{geom}} = \Delta \phi_w .
\label{eq:fgeom}
\end{equation}

The geometric coupling factor $f_{\mathrm{geom}}$ encodes the detector geometry and readout configuration through the weighting potential. For the double gap geometry considered here, it represents the combined contribution of both gas gaps and is constant for a fixed detector layout.

At the macroscopic level, the total induced charge produced by an event is proportional to the number of primary electron ion pairs and to the amplification of the primary charge under the applied electric field. This allows the induced charge to be expressed as

\begin{equation}
Q_{\mathrm{ind}}(E) = e\, f_{\mathrm{geom}}\, N_0(E)\, G(E),
\label{eq:Qind_geom}
\end{equation}

where $N_0(E)$ is the primary ionization yield obtained from Eq.~(\ref{eq:N0_from_Edep}), and $G(E)$ is an effective, field dependent amplification factor. Equation~(\ref{eq:Qind_geom}) introduces the first explicit source of field dependence in the detector response. The efficiency turn on therefore originates from the field dependence of the effective amplification term $G(E)$, as the induced charge exceeds the detection threshold with increasing electric field.

\subsection{Gain reconstruction} \label{subsec:gainRec}

Equation~(\ref{eq:Qind_geom}) provides a direct link between the measurable detector sensitivity and an effective amplification factor. Since the primary ionization yield $N_0(E)$ is essentially independent of the electric field, any field dependence of the induced charge must be encoded in the amplification term $G(E)$. Within this macroscopic framework, the gain is defined as an effective amplification relating the primary ionization yield to the induced charge on the readout electrodes. From Eq.~(\ref{eq:Qind_geom}), the gain can be written as

\begin{equation}
G(E) = \frac{Q_{\mathrm{ind}}(E)}
{e\, f_{\mathrm{geom}}\, N_0(E)} ,
\label{eq:G_def}
\end{equation}

where $Q_{\mathrm{ind}}(E)$ is the induced charge inferred from the detector sensitivity, $f_{\mathrm{geom}}$ is the geometric coupling factor, and $N_0(E)$ is the primary ionization yield obtained from Geant4.

For a fixed muon beam energy and detector geometry, this definition does not provide an absolute gain scale, as microscopic avalanche development and gas transport properties are not modeled. The gain must therefore be normalized using experimental information. In this work, the standard CMS gas mixture (STD), for which validated efficiency measurements under GIF++ conditions are available, is used as the experimental reference.

\subsubsection{Experimental and Effective Gain definitions}\label{subsubsec:ExpEffGain}

For the standard CMS gas mixture (STD), the reconstructed gain corresponds to an \textbf{experimental gain}. This gain is anchored to validated efficiency measurements obtained at GIF++ under controlled irradiation conditions~\cite{gomes2024performance}. As such, it provides the absolute normalization of the amplification scale used in this work.

For the remaining gas mixtures, the gain reconstructed from Eq.~(\ref{eq:G_def}) is defined as an \textbf{effective gain}. These effective gains are not absolute quantities, but are obtained relative to the STD reference by combining the reconstructed induced charge with the corresponding primary ionization yield. 

This procedure allows the response of different mixtures to be compared within a unified macroscopic framework, while preserving the experimentally validated normalization provided by the STD mixture.

\subsubsection{Extraction of effective Townsend parameters}
\label{subsubsec:TownsendCoef}

From the gain reconstructed in Eq.~(\ref{eq:G_def}), an effective Townsend coefficient is obtained by assuming the usual exponential amplification across a single gas gap,

\begin{equation}
\alpha_{\mathrm{eff}}(E)
= \frac{1}{d_{\mathrm{gap}}}\,\ln G(E),
\label{eq:alpha_from_G}
\end{equation}

with $d_{\mathrm{gap}}$ the gas-gap thickness. The field dependence of $\alpha_{\mathrm{eff}}(E)$ is then described with

\begin{equation}
\alpha_{\mathrm{eff}}(E)
= A\,p\,\exp\!\left(-\frac{B\,p}{E}\right),
\label{eq:alpha_param}
\end{equation}

where $p$ is the gas pressure and $(A,B)$ are macroscopic parameters. Taking the logarithm gives a linear relation,

\begin{equation}
\ln\!\left( \frac{\alpha_{\mathrm{eff}}(E)}{p} \right)
= \ln A - B\,\frac{p}{E},
\label{eq:AB_linear}
\end{equation}

so that $(A,B)$ are extracted from a linear fit of $\ln(\alpha_{\mathrm{eff}}/p)$ versus $1/E$.

With the effective gain $G(E)$ reconstructed and parametrized through the coefficients $(A,B)$, the detector response can be propagated to any gas mixture within the same macroscopic framework. For each mixture, the corresponding primary ionization yield $N_0(E)$ obtained from Geant4 is combined with the reconstructed amplification to compute the induced charge, the detector sensitivity, and ultimately the efficiency as a function of the applied voltage. 
 
In this way, the experimentally anchored STD mixture provides the reference scale, while the remaining mixtures yield effective sensitivities and efficiency curves derived consistently from the same reconstruction procedure. The full reconstruction sequence implemented in this work is summarized in Fig.~\ref{fig:pipeline}.

The reconstruction sequence is:

\begin{figure}[htbp]
\centering
\resizebox{\linewidth}{!}{%
\begin{tikzpicture}[
    node distance=1.3cm,
    every node/.style={font=\small},
    box/.style={
        rectangle, draw=black, rounded corners,
        minimum height=0.9cm,
        minimum width=3.0cm,
        align=center
    },
    arr/.style={->, draw=black, line width=0.9pt}
]

\node[box] (expEff) {GIF++ / ICHEP\\ $\varepsilon_{\mathrm{STD}}(HV)$};
\node[box, right=1.3cm of expEff] (stdS) {$S_{\mathrm{STD}}(E)$};
\node[box, right=1.3cm of stdS] (stdGA)
  {$G_{\mathrm{STD}}(E)$\\ $(A_{\mathrm{STD}},B_{\mathrm{STD}})$};

\node[box, below=1.15cm of expEff] (g4N0)
  {Geant4\\ $\langle N_0^{\mathrm{mix}}(E)\rangle$\\ Eq.~(3.2)};

\node[box, right=1.55cm of g4N0] (mixGA)
  {$G_{\mathrm{mix}}(E)$\\ $(A_{\mathrm{mix}},B_{\mathrm{mix}})$};

\node[box, right=1.3cm of mixGA] (Smix) {$S_{\mathrm{mix}}(E)$};
\node[box, right=1.3cm of Smix] (EffHV) {$\varepsilon_{\mathrm{mix}}(HV)$};

\draw[arr] (expEff) -- (stdS);
\draw[arr] (stdS)   -- (stdGA);

\draw[arr] (g4N0) -- (mixGA);

\draw[arr] (mixGA) -- (Smix);
\draw[arr] (Smix)  -- (EffHV);

\path (stdGA.south) ++(0,-0.6) coordinate (pDown);     
\coordinate (pLeft) at (mixGA.north |- pDown);         
\draw[arr] (stdGA.south) -- (pDown) -- (pLeft) -- (mixGA.north);

\end{tikzpicture}%
}
\caption{Macroscopic reconstruction pipeline. The experimental efficiency of the STD mixture at GIF++ is converted into a field-dependent sensitivity $S_{\mathrm{STD}}(E)$ and then into the absolute gain and macroscopic parameters $G_{\mathrm{STD}}(E)$ and $(A_{\mathrm{STD}},B_{\mathrm{STD}})$. Combined with the Geant4 primary ionization yields $\langle N_0^{\mathrm{mix}}(E)\rangle$, the framework reconstructs $G_{\mathrm{mix}}(E)$, $S_{\mathrm{mix}}(E)$ and the efficiency curves $\varepsilon_{\mathrm{mix}}(HV)$.}
\label{fig:pipeline}
\end{figure}
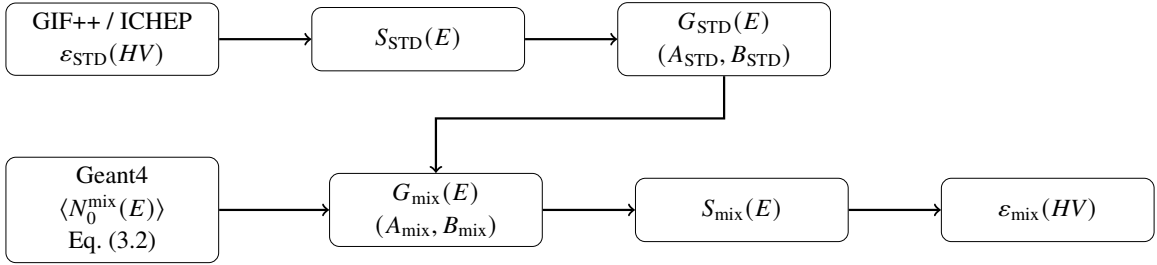

This macroscopic framework provides the first predictive estimates of ECO1 and ECO2 performance in an iRPC geometry without requiring gas-transport data that are not currently available for HFO mixtures.
\section{Results} \label{sec:results}
This section presents the results obtained from the macroscopic reconstruction procedure described in Section~\ref{sec:Methodology}. The reconstruction of the primary ionization yield requires the effective ionization energy of each gas mixture, which enters exclusively through Eq.~(\ref{eq:N0_from_Edep}) and is computed using the weighted average defined in Eq.~(\ref{eq:Wmix}). The effective $W_{\mathrm{mix}}$ values used throughout this work are summarized in Table~\ref{tab:Wvalues_results}.

\begin{table}[htbp]
\centering
\caption{Effective ionization energies $W_{\mathrm{mix}}$ used for the calculation of the primary ionization yield for each gas mixture.}
\label{tab:Wvalues_results}
\smallskip
\small
\begin{tabular}{l c}
\hline
Gas mixture & $W_{\mathrm{mix}}$ [eV] \\
\hline
STD      & 34.20 \\
Mix I    & 34.15 \\
Mix II   & 34.12 \\
ECO Mix I & 34.40 \\
ECO Mix II& 34.05 \\
\hline
\end{tabular}
\end{table}

The effective ionization energies vary only marginally among mixtures and affect the reconstruction only through the primary ionization yield. No field dependence is introduced. The Geant4 primary ionization yield is field independent and nearly identical for all mixtures, confirming that it cannot drive the efficiency turn-on (Appendix~\ref{app:PrimaryIonization}).
\subsection{Absolute Gain calibration for the STD mixture}
\label{subsec:results_STD_gain}

The absolute amplification scale of the macroscopic model is fixed using the standard CMS gas mixture (STD), for which validated efficiency measurements at GIF++ are available.  

\begin{figure}[htbp]
\centering
\includegraphics[width=.45\textwidth]{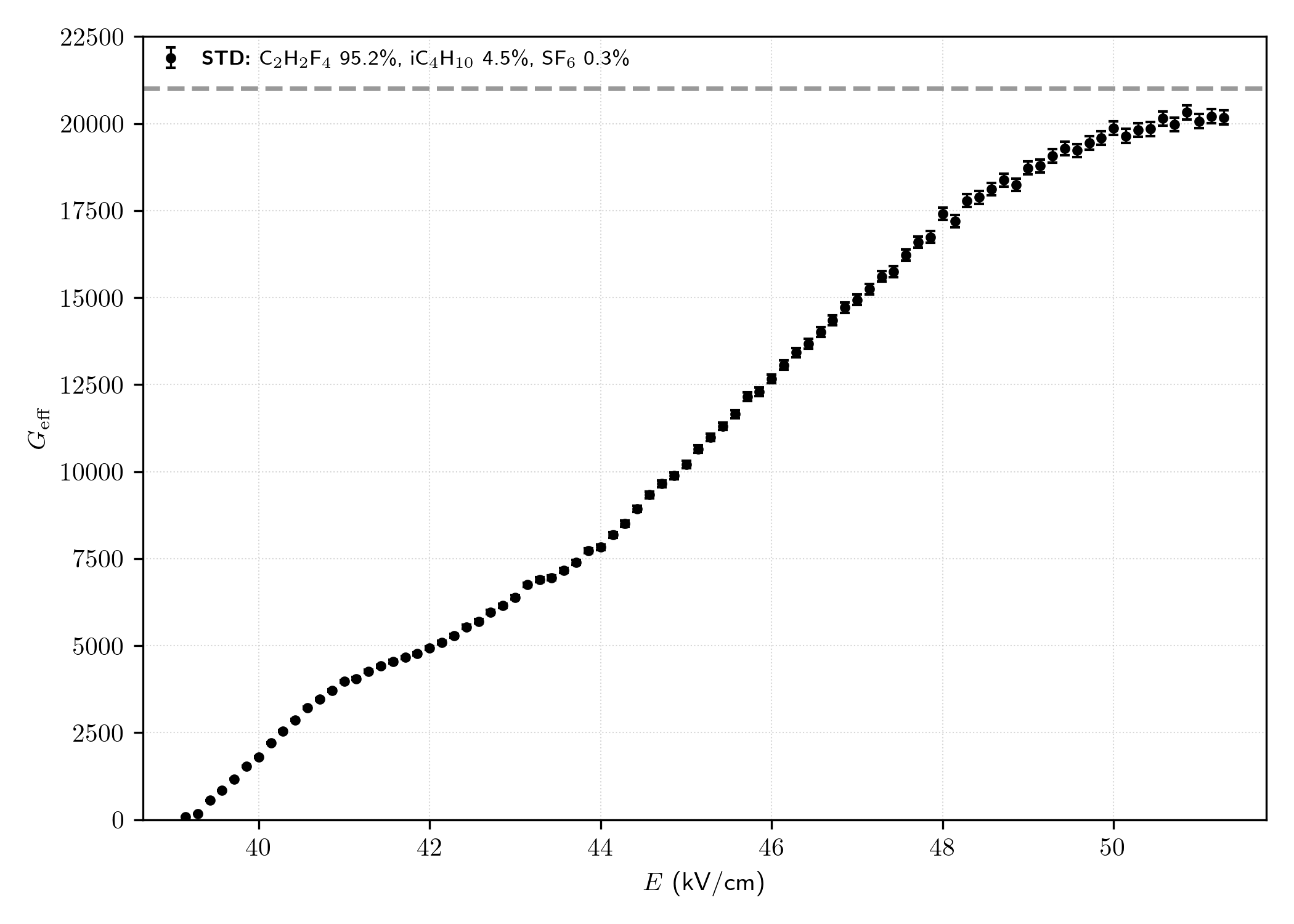}
\qquad
\includegraphics[width=.45\textwidth]{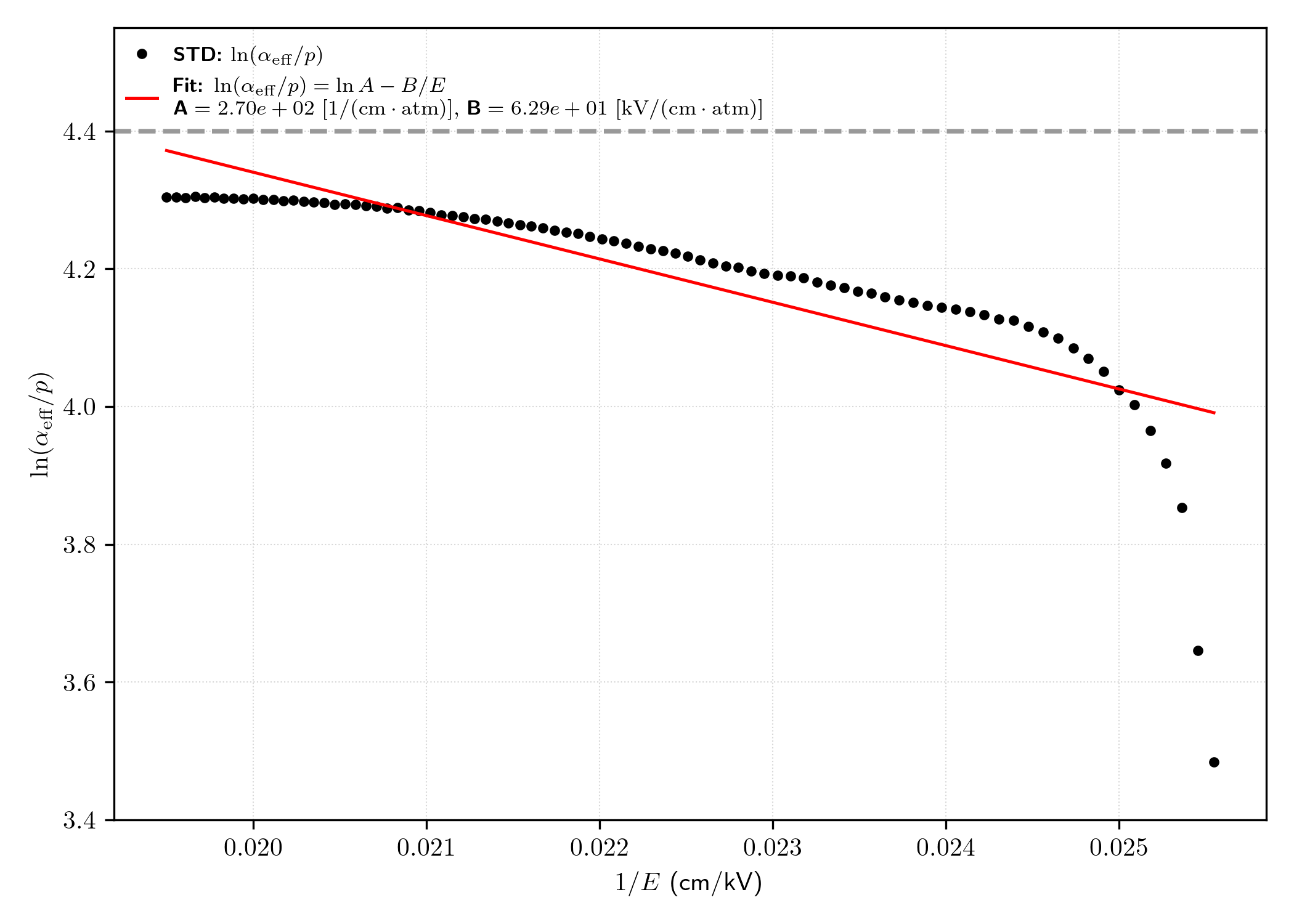}
\caption{Reconstructed effective gain $G_{\mathrm{STD}}(E)$ for the standard CMS gas mixture, obtained from the experimental efficiency curve measured at GIF++ (left). Townsend Fit representation for the STD mixture (right).\label{fig:ExpGanABCoef}}
\end{figure}

Figure~\ref{fig:ExpGanABCoef} (left) shows the reconstructed effective gain $G_{\mathrm{STD}}(E)$ as a function of the electric field. The gain displays the characteristic avalanche behavior expected for RPC operation, with a steep rise over a narrow field interval followed by a broad high-gain region. This curve defines the absolute normalization of the gain used throughout this work.

The same reconstructed gain is converted into an effective Townsend representation, shown in Figure~\ref{fig:ExpGanABCoef} (right). The linear dependence of $\ln(\alpha_{\mathrm{eff}}/p)$ on $1/E$ over a wide field range validates the macroscopic Townsend parametrization for the reference mixture and allows the extraction of the parameters $(A_{\mathrm{STD}}, B_{\mathrm{STD}})$.

These parameters provide the experimental anchor for the reconstruction of the effective gain and detector response of all other gas mixtures.
\subsection{Effective gain reconstruction for all mixtures}
\label{subsec:results_relative_gain}

The effective gain $G_{\mathrm{eff}}(E)$ is reconstructed for all gas mixtures. The corresponding macroscopic Townsend parameters and working points are summarized in Table~\ref{tab:ABvalues}.

\begin{table}[htbp] 
\centering 
\caption{Effective macroscopic Townsend parameters $(A,B)$, relative primary ionization, and working point values for the five gas mixtures.} 
\label{tab:ABvalues} 
\smallskip
\small
\begin{tabular}{l|ccccc} 
\hline 
Mixture & $A~[\mathrm{cm^{-1}\,atm^{-1}}]$ & $B~[\mathrm{kV/(cm\,atm)}]$ &
$N_0^{\mathrm{mix}}/N_0^{\mathrm{STD}}$ & $HV_{\mathrm{WP}}~[\mathrm{V}]$
& $E_{\mathrm{WP}}~[\mathrm{kV/cm}]$ \\
\hline 
STD     & 270.42 & 62.93 & 1.000 & 7220 & 51.57 \\ 
Mix1    & 275.31 & 61.01 & 1.019 & 7000 & 50.00 \\ 
Mix2    & 275.73 & 60.58 & 1.021 & 6950 & 49.64 \\ 
ECO1    & 270.08 & 71.47 & 1.000 & 8200 & 58.57 \\ 
ECO2    & 270.08 & 67.11 & 1.000 & 7700 & 55.00 \\ 
\hline
\end{tabular} 
\end{table}

\begin{figure}[htbp]
\centering
\includegraphics[width=.4\textwidth]{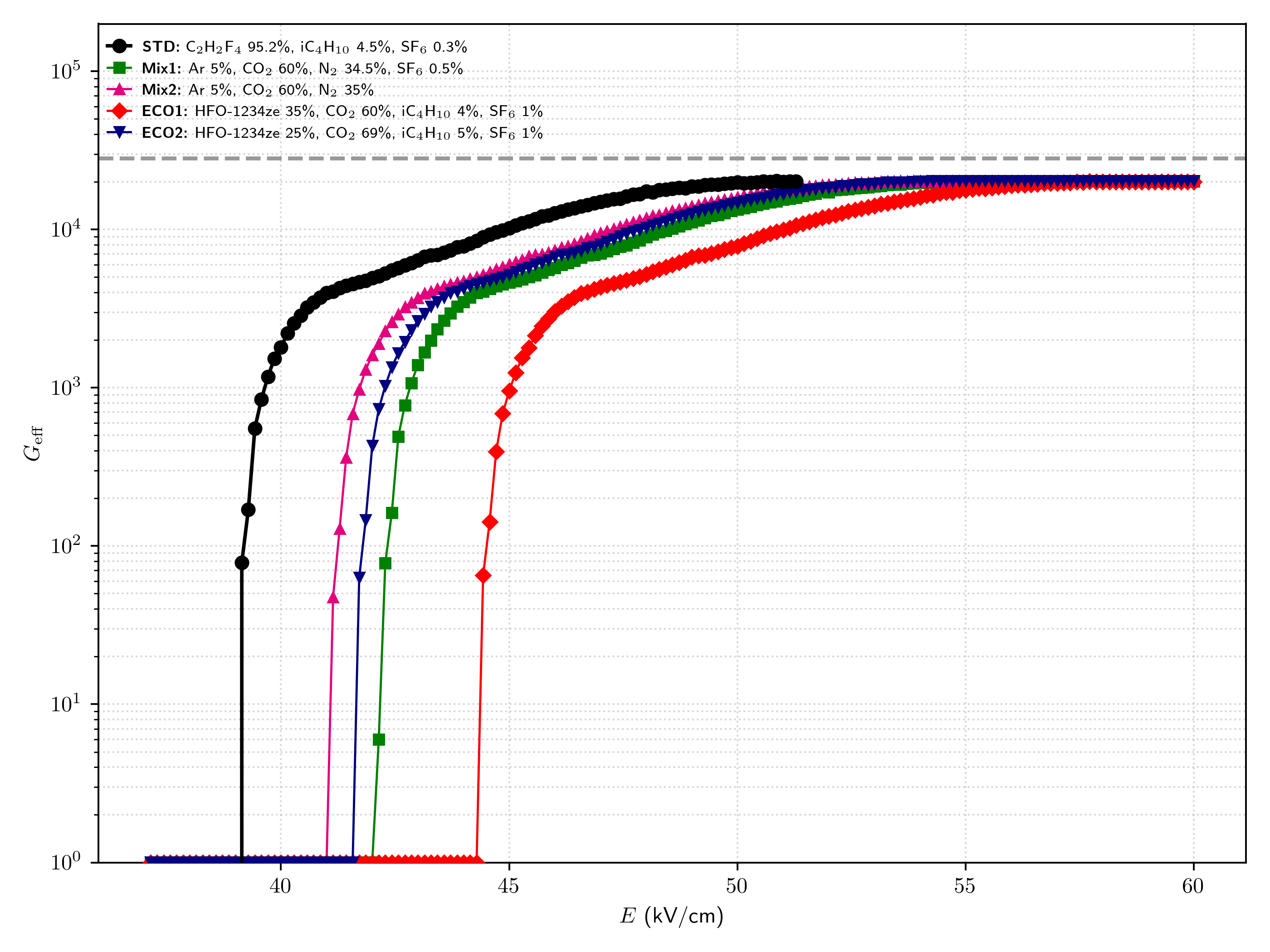}
\caption{Reconstructed effective gain $G_{\mathrm{eff}}(E)$ for all gas mixtures.}
\label{fig:GeffMixes}
\end{figure}

Figure~\ref{fig:GeffMixes} shows the reconstructed effective gain as a function of the electric field for all mixtures. The STD mixture defines the absolute scale, while the remaining mixtures inherit their normalization through the reconstructed macroscopic parameters.

Mix1 and Mix2 closely follow the STD behavior, reaching comparable gain values at lower fields. In contrast, ECO1 and ECO2 exhibit a delayed gain onset and require higher operating fields to achieve the same effective amplification, in agreement with their larger $B$ parameters and higher working point voltages.
\subsection{Sensitivity and Efficiency}
\label{subsec:results_sensitivity_efficiency}

\begin{figure}[htbp]
\centering
\includegraphics[width=.45\textwidth]{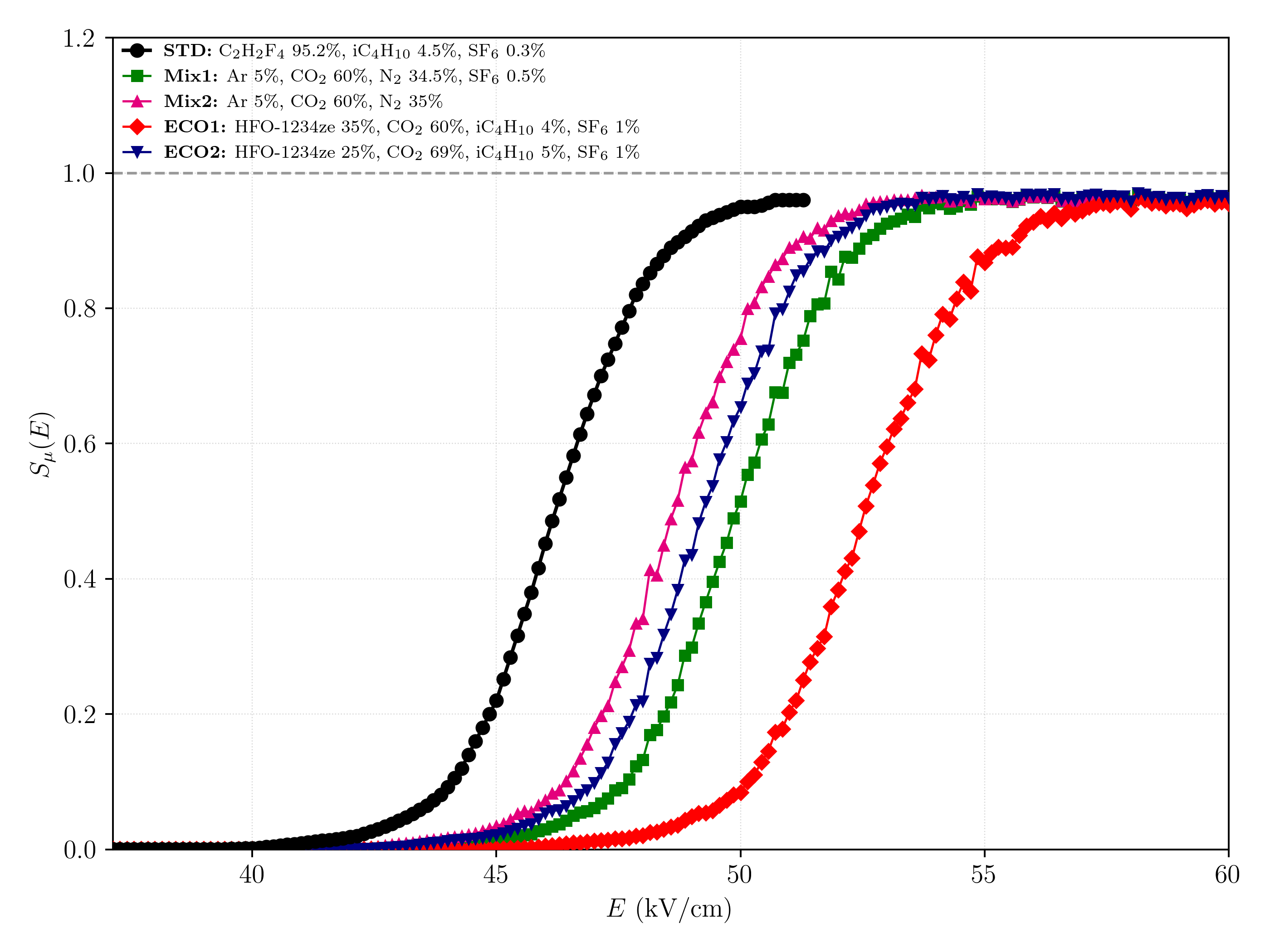}
\hfill
\includegraphics[width=.45\textwidth]{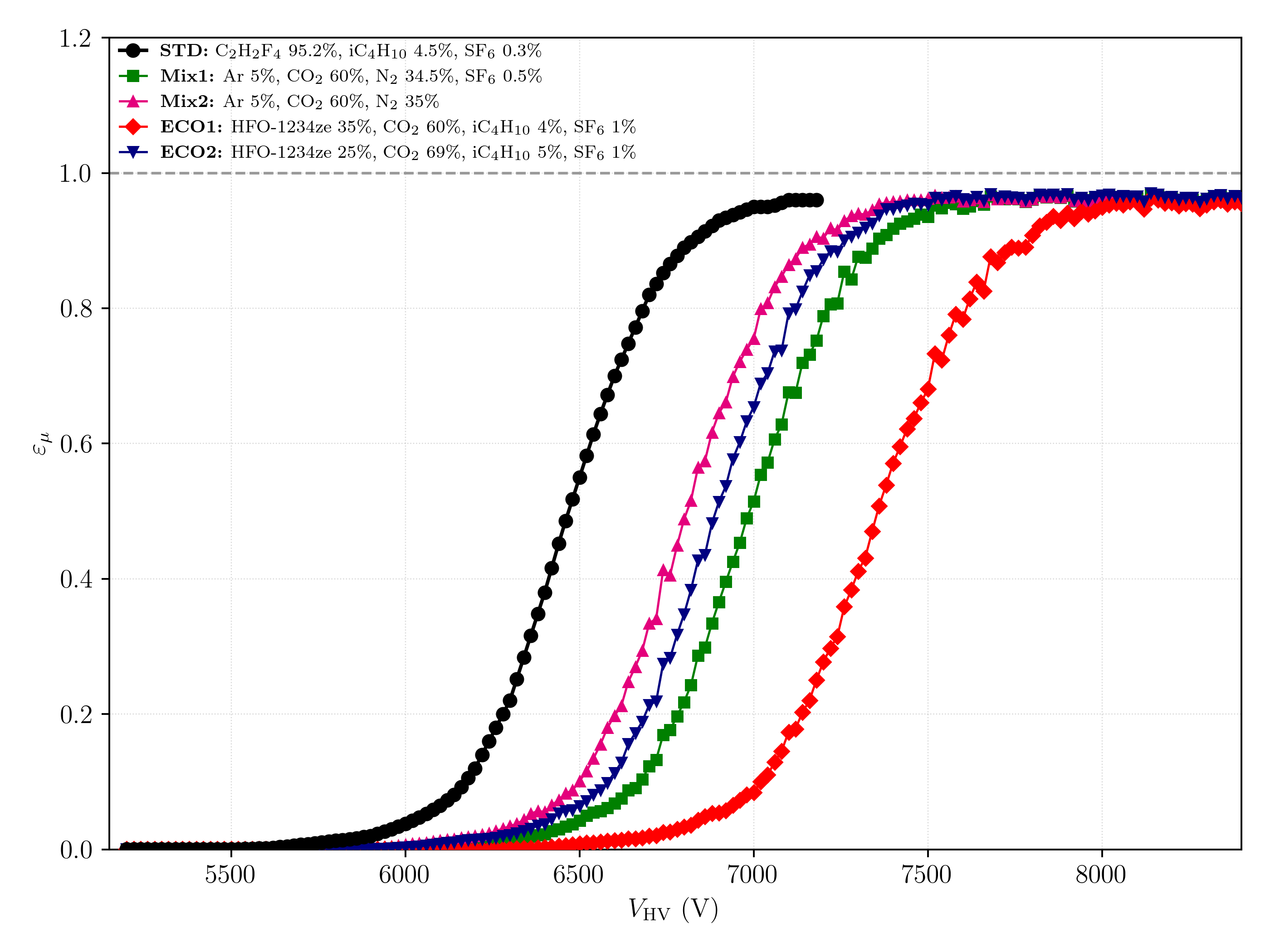}
\caption{
Reconstructed muon response for all gas mixtures.
(left) Sensitivity $S_\mu(E)$ as a function of the effective electric field.
(right) Efficiency $\varepsilon_\mu(HV)$ as a function of the applied high voltage.
In both panels, the dashed horizontal line marks the 95\% level used to define the working point.
}
\label{fig:SensitivityEfficiencyMixes}
\end{figure}

Figure~\ref{fig:SensitivityEfficiencyMixes} summarizes the reconstructed muon response for all gas mixtures in terms of sensitivity and efficiency. The left panel shows the sensitivity $S_\mu(E)$ as a function of the effective electric field, while the right panel presents the corresponding efficiency curves $\varepsilon_\mu(HV)$ obtained through the field--voltage mapping.

In all cases, the response exhibits the characteristic sigmoid shape expected for RPC operation, with a rapid transition between a low-response region and a high-efficiency plateau. The standard CMS mixture (STD) reaches the 95\% sensitivity level at $E \simeq 51.5$~kV/cm, corresponding to a working point around $HV_{\mathrm{WP}}\simeq 7.2$~kV. The CO$_2$-based mixtures Mix1 and Mix2 follow closely, with working points shifted by less than $\sim 300$~V.

The eco-friendly mixtures display a systematic displacement of the curves toward higher fields and voltages. ECO2 reaches the 95\% efficiency level at approximately $HV_{\mathrm{WP}}\simeq 7.7$~kV, while ECO1 exhibits the latest turn-on, with a working point around $8.2$~kV. The broader transition region observed for the eco-friendly mixtures reflects their reduced effective gain at fixed field.

\section{Model validation} \label{sec:model_validation}

Table~\ref{tab:WP_validation} compares the working points obtained with the macroscopic reconstruction to values reported for iRPC prototype measurements. All working points are quoted at the 95\% efficiency level. In the external references, gas mixture~II and gas mixture~IV correspond to Mix1 and Mix2 in this work, respectively~\cite{supratik2025perspectives}. For the eco-friendly mixtures, our ECO1 (ECO2 ~\cite{gomes2024performance}) and our ECO2 (ECO3 in ~\cite{gomes2024performance}) are matched accordingly.

\begin{table}[htbp]
\centering
\caption{Working-point comparison at 95\% efficiency. Literature values are taken from Refs.~\cite{supratik2025perspectives,gomes2024performance} using the mixture mapping described in the text.}
\label{tab:WP_validation}
\small
\setlength{\tabcolsep}{5pt}
\begin{tabular}{lcc}
\hline
Mixture & $HV_{\mathrm{WP}}$ [V] & $HV_{\mathrm{WP}}^{\mathrm{ref}}$ [V] \\
\hline
STD  & 7220 & 7220 \\
Mix1 & 7000 & 11900 \\
Mix2 & 6950 &  8800 \\
ECO1 & 8200 & 7700 \\
ECO2 & 7700 & 8200 \\
\hline
\end{tabular}
\end{table}

In both the reconstruction and the reported measurements, the CO$_2$ mixtures (Mix1 and Mix2) operate close to the STD reference, whereas the HFO-based mixtures require higher working voltages to reach the same target efficiency.

\subsection*{Consistency with reconstructed Townsend parameters}

The observed working point are consistent with the reconstructed macroscopic Townsend parameters. Mixtures exhibiting larger $B$ values require higher electric fields to reach the same effective gain, leading to delayed turn-on and increased working voltages. This relation between the effective Townsend parameters $(A_{\mathrm{mix}}, B_{\mathrm{mix}})$ and $HV_{\mathrm{WP}}$ provides an internal consistency check of the reconstruction.
\section{Limitations of the macroscopic approach} \label{Sec:Limitations}

The reconstruction presented is macroscopic and it does not aim to describe the microscopic gas physics inside the RPC. Several intrinsic limitations must therefore be acknowledged.

\begin{itemize}
    \item All field dependence is encoded in an effective gain and an effective Townsend coefficient, which cannot capture fluctuations, space charge effects, or streamer formation.
    \item The reconstructed Townsend parameters $(A,B)$ should not be interpreted as gas properties. They represent lumped macroscopic quantities that combine multiple physical mechanisms, including ionization, attachment, quenching and detector geometry.
\end{itemize}

A complementary discussion, together with the reconstructed $\alpha_{\mathrm{eff}}(E)$ curves, is provided in Appendix~\ref{app:Limitation}. Figure~\ref{fig:alphaMixes} shows the Townsend representation $\ln(\alpha_{\mathrm{eff}}/p)$ as a function of $1/E$ for the five gas mixtures. In the exponential growth region, all mixtures exhibit an approximately linear behavior, justifying the use of an effective Townsend parametrization. Deviations from linearity at high $1/E$ reflect the breakdown of the effective description outside the fitted gain region rather than genuine gas effects.
\section{Conclusions} \label{sec:conclusions}

The main outcome of this work is the demonstration that detector performance indicators, such as effective gain and working point, can be reconstructed within a fully macroscopic framework, without requiring explicit knowledge of the microscopic gas dynamics. This enables a direct and consistent comparison of gas mixtures based solely on their observable detector response.

The reconstructed working points and their relative ordering among mixtures are found to be consistent with published iRPC measurements, providing confidence in the internal coherence of the approach and in its ability to capture the dominant field dependent behavior governing detector operation.

Beyond reproducing known results, the relevance of this framework lies in its predictive capability. It allows the exploration and ranking of alternative gas mixtures, including eco-friendly candidates, using simulation driven inputs only. This is particularly important in regimes where microscopic tools are not yet available or validated, as is currently the case for several HFO mixtures.

Overall, the method provides a practical tool to estimate working points, relative gain behavior, and operating margins prior to experimental testing. As such, it offers a complementary strategy to guide mixture selection and detector optimization for future iRPC developments, reducing experimental overhead and accelerating the evaluation of new gas compositions.
\appendix
\section{Field independence of the primary ionization yield}
\label{app:PrimaryIonization}

\begin{figure}[htbp]
\centering
\includegraphics[width=.4\textwidth]{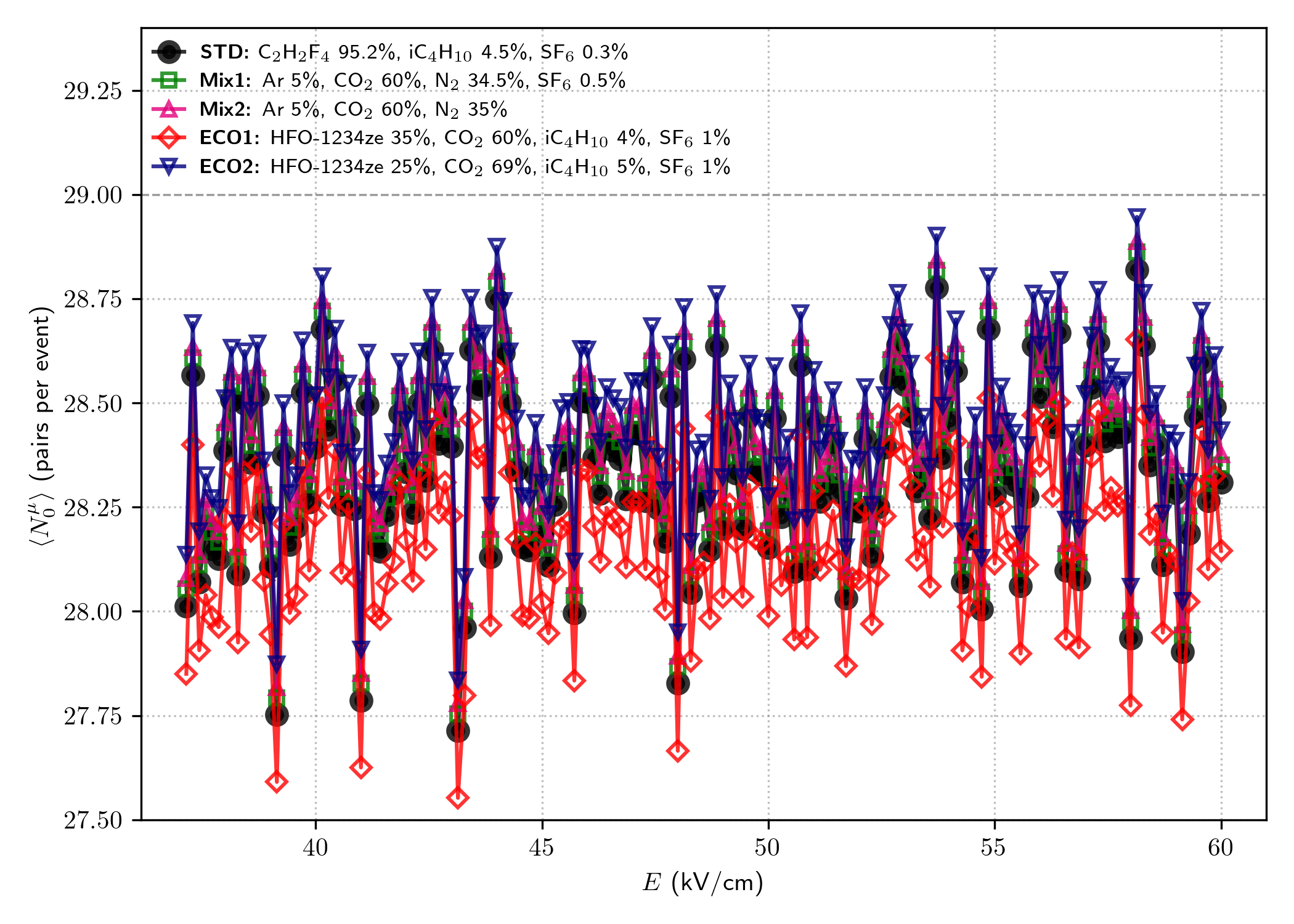}
\caption{Mean primary ionization $\langle N_0^{\mu}(E)\rangle$ for all gas mixtures
as a function of the effective field.}
\label{fig:N0mu}
\end{figure}

Figure~\ref{fig:N0mu} shows the mean number of primary ionization pairs $\langle N_0^{\mu}(E)\rangle$ produced by 6~GeV muons in the double gap iRPC, using Eq.~(\ref{eq:N0_from_Edep}). os a hacer
osahacerWithin statistical uncertainties, $\langle N_0^{\mu}(E)\rangle$ is constant over the full explored field range. Differences among gas mixtures are small and consistent with their respective densities and effective $W$ values. No systematic dependence on the electric field is observed. This result confirms that, within a pure Geant4 description, the efficiency turn on cannot originate from primary ionization and must be driven by the effective gain $G(E)$.

\section{Effective Townsend parametrization and limitations} \label{app:Limitation}

\begin{figure}[htbp]
\centering
\includegraphics[width=.4\textwidth]{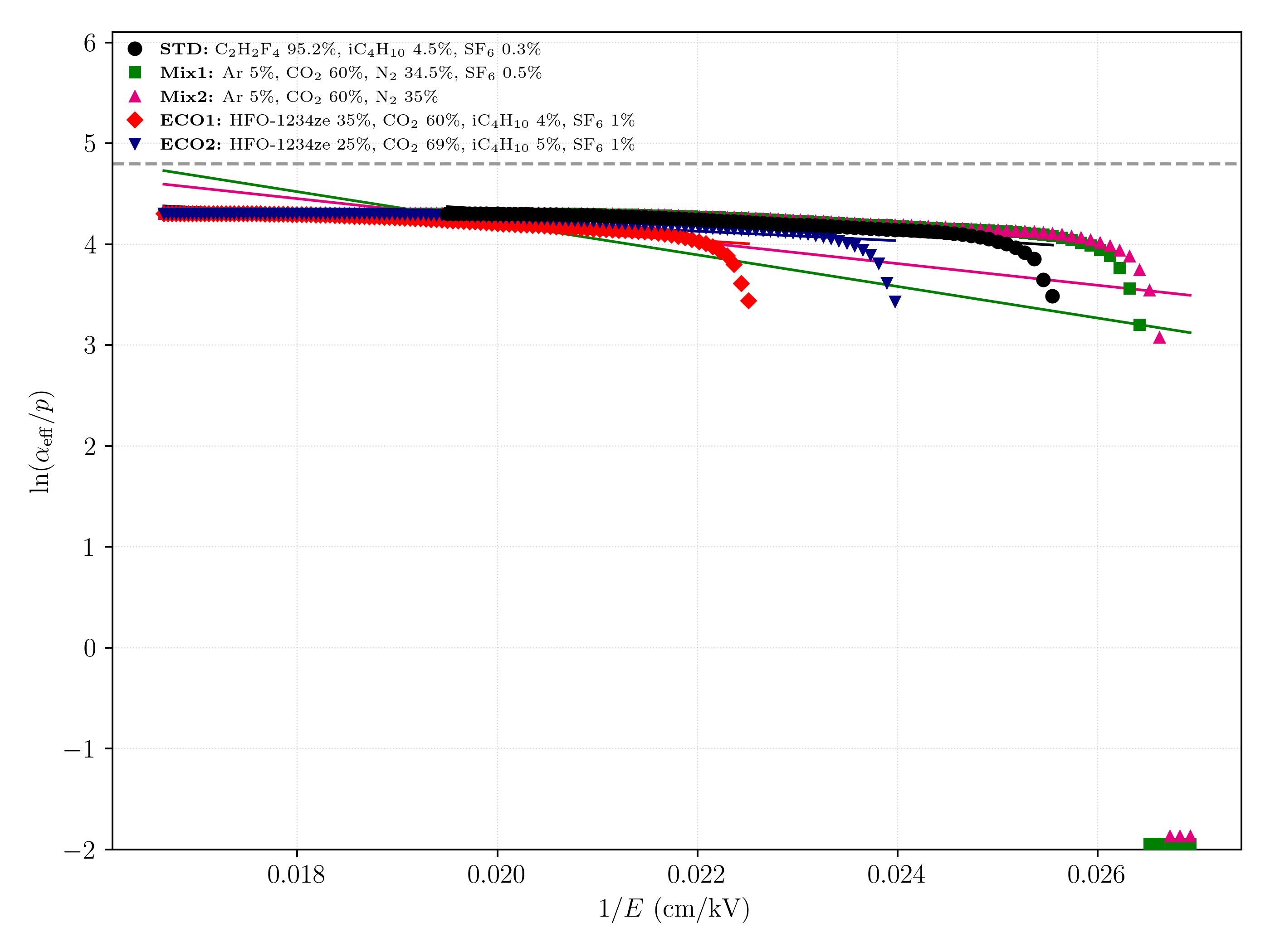}
\caption{Effective Townsend representation $\ln(\alpha_{\mathrm{eff}}/p)$ as a
function of $1/E$ for the five gas mixtures. Solid lines correspond to linear
fits used to extract the macroscopic parameters $(A,B)$.}
\label{fig:alphaMixes}
\end{figure}

Figure~\ref{fig:alphaMixes} shows the reconstructed effective Townsend coefficient $\alpha_{\mathrm{eff}}(E)$ for all gas mixtures. Deviations at low and high fields reflect the limited validity of the effective description outside the exponential avalanche regime and illustrate the intrinsic limitations of a macroscopic approach.

\nocite{*}
\bibliographystyle{JHEP}
\bibliography{biblio.bib}

\end{document}